\documentstyle[11pt,paspconf,epsf]{article}

\def\Lya{Ly$\alpha\ $}
\def\HI{\hbox{H~$\scriptstyle\rm I\ $}}
\def\HII{\hbox{H~$\scriptstyle\rm II\ $}}
\def\nH{{\rm H}}
\def\nHI{{\rm HI}}
\def\etal{{\it et al.\ }}

\begin{document}

\title{The Dark Ages:\ Searching for the Epoch of First Light\\}
\author{A Meiksin}
\affil{Institute for Astronomy, University of Edinburgh, Royal Observatory,
Edinburgh\ EH9\ 3HJ}

\begin{abstract}
Prior to the epoch of full reionization, the intergalactic medium and
gravitationally collapsed systems will be detectable in 21-cm radiation.
Physical mechanisms that would produce a 21-cm signature are discussed.
These include Lya coupling of the hydrogen spin temperature to the kinetic
temperature of the gas resulting from the radiation by an early generation
of stars, preheating by soft x-rays from collapsing dark matter halos, and
preheating by ambient Lya photons. A patchwork of either 21-cm emission, or
absorption against the Cosmic Microwave Background, will result. Large
radio telescopes like the Giant Metre Wavelength Telescope (GMRT) or a Square
Kilometre Array Radio Telescope (SKA) offer the prospect of measuring
this signature, and so detecting the transitional epoch from a dark universe to
one with light.\\
\\
{\it Invited review talk at the International SKA Science Meeting, Calgary,
19--22 July 1998. Included in the SKA Radio Telescope Science Case, ed. R. Braun.}
\end{abstract}

\section{Introduction}
The development of structure in the Universe was well advanced at
early times.  Quasars have been detected nearly to a redshift of
$z=5$, and the most distant galaxies to even greater redshifts (Dey
\etal 1998). The spectra of high redshift QSOs have additionally shown
that the Intergalactic Medium (IGM) itself had undergone an extensive
development of nonlinear structures at early times as well, as
revealed by the \Lya forest. Still unknown, however, is the nature of
energetic processes at these early times. While numerical simulations
have shown that the IGM is expected to fragment into structures at
early times in Cold Dark Matter (CDM) dominated cosmologies (Zhang
\etal 1998), and even into early galaxies (Governato \etal 1997), the
simulations are much less able to predict the efficiency with which
gravitationally collapsed objects will emit radiation.  Although QSO
sources may account for the photoionizing UV background at high
redshifts (Meiksin \& Madau 1993; Haardt \& Madau 1997), it is less
clear that they were responsible for the original reionization of the
IGM.  Similarly, although $z>5$ galaxies have been detected, the epoch
during which the first generation of stars formed is still poorly
constrained (Madau \etal 1998; Hughes \etal 1998).  Although IR
observations will permit even higher redshift galaxies and QSOs to be
observed, detections become increasingly difficult because of the
diminution in surface brightness due to cosmological expansion. This
difficulty calls for alternative means to be found for discovering the
nature and period of the first major generation of energy-producing
sources, the epoch of First Light.

I discuss here possible ways in which a Square Kilometre Array (SKA) radio
telescope could reveal the first epochs of energy generation. Because
so little is known about the distribution and nature of such sources,
the emphasis will be on the physical mechanisms related to their
detection. The means by which the sources are revealed is through their
impact on the surrounding neutral IGM and the resulting emission or absorption
of 21-cm radiation. In the next section, I review the Wouthuysen-Field
mechanism for the excitation of 21-cm emission through the scattering of
\Lya photons, and the heating of the neutral IGM by resonant scattering of
the \Lya photons and soft x-ray heating from early collapsed objects.
In \S III, a few scenarios are presented for the possible effects of early
generations of radiation sources on the IGM and their detection.

\section{21-cm Emission and Absorption Mechanisms}

\subsection{The Spin Temperature}

The emission or absorption of 21-cm radiation from a neutral IGM is
governed by the spin temperature $T_S$ of the hydrogen, defined by
\begin{equation}
\frac{n_1}{n_0}=3\exp[-T_*/ T_S],
\end{equation}
where $n_0$ and $n_1$ are the singlet and triplet $n=1$ hyperfine
levels, $T_*\equiv h\nu_{10}/k_{\rm B}=0.07\,$K, where $\nu_{10}$ is the
frequency of the 21-cm transition, $h$ is Planck's constant, and $k_{\rm B}$
is Boltzmann's constant. In the presence of only the Cosmic Microwave
Background (CMB) radiation, the spin temperature will be the same as the
temperature of the CMB, and no emission or absorption relative to the CMB will
be detectable. A mechanism is required that decouples the two
temperatures. This may be achieved by coupling the spin temperature to
the kinetic temperature of the gas itself. Two mechanisms are
available, collisions between hydrogen atoms (Purcell \& Field 1956)
and scattering by \Lya photons (Wouthuysen 1952; Field 1958). The
collision-induced coupling between the spin and kinetic temperatures
is dominated by the spin-exchange process between the colliding
hydrogen atoms. The rate, however, is too small for realistic IGM
densities at the redshifts of interest, although collisions may be
important in dense regions,
$\delta\rho/\rho\ga30[(1+z)/10]^{-2}$ (Madau, Meiksin, \& Rees 1997).

Instead the dominant mechanism is likely to be \Lya scattering through
the Wouthuysen-Field effect. This process mixes the hyperfine levels of
neutral hydrogen in its ground state via an intermediate transition to
the $2p$ state. An atom initially in the $n=1$ singlet state may absorb
a \Lya photon that puts it in an $n=2$ state, allowing it to return to the
triplet $n=1$ state by a spontaneous decay. At this point, the astute student
of quantum mechanics will ask how is it possible for electric dipole radiation
(\Lya photons) to induce a spin transition? The key is spin-orbit coupling:\
it's the total angular momentum $F=I+J$ that counts. (Here $I$ is the proton
spin and $J$ is the total electron angular momentum, $J=S+L$.) There are
four hyperfine states involved, the $n=1$ singlet $_0S_{1/2}$ and triplet
$_1S_{1/2}$ states (the notation is $_FL_J$), and the two triplet $n=2$ states
$_1P_{1/2}$ and $_1P_{3/2}$. The selection rule $\Delta F=0,1$ permits the
transitions $_0S_{1/2}\rightarrow\, _1P_{1/2},\, _1P_{3/2}$ and
$_1P_{1/2},\, _1P_{3/2}\rightarrow\, _1S_{1/2}$, and so effectively
$_0S_{1/2}\rightarrow\, _1S_{1/2}$ occurs via one of the $n=2$ states.

When the IGM is highly opaque to the scattering of \Lya photons, as it is when
still neutral, the large number of scatterings of \Lya photons in an ambient
radiation field will ensure a Boltzmann distribution for the photon energies
near the \Lya frequency, with a temperature given by the kinetic temperature
$T_K$ of the IGM (Field 1959). In this case, the spin temperature of the
neutral hydrogen becomes\footnote{In the presence of a radio source,
the antenna temperature of the radio emission should be added to $T_{\rm CMB}$
in equation~\ref{eq:Tspin}. The radio emission may make an important
contribution in the vicinity of a radio-loud quasar (Bahcall \& Ekers 1969),
and would itself permit the IGM to be detected in 21-cm radiation.}
\begin{equation}
T_S=\frac{T_{\rm CMB}+y_\alpha T_K}{1+y_\alpha},  \label{eq:Tspin}
\end{equation}
where $T_{\rm CMB}=2.73(1+z)$~K is the temperature of the CMB (Mather \etal
1994), and
\begin{equation}
y_\alpha\equiv\frac{P_{10}}{A_{10}}\frac{T_*}{T_K}
\end{equation}
is the \Lya pumping efficiency. Here, $A_{10}=2.9\times 10^{-15}\,$s$^{-1}$
is the spontaneous decay rate of the hyperfine transition of atomic hydrogen,
$P_{10}$ is the indirect de-excitation rate of the triplet via absorption of a
\Lya photon to the $n=2$ level, and $T_S\gg T_*$ was assumed. In the
absence of \Lya pumping the spin temperature goes to equilibrium with the
21-cm background radiation field on a timescale $T_*/(T_{\rm CMB}A_{10})\approx
5\times10^4\,$yr, and neutral intergalactic hydrogen will produce neither an
absorption nor emission signature. If $y_\alpha$ is large, $T_S\rightarrow
T_K$, signifying equilibrium with the matter. A consideration
of the net transition rates between the various hyperfine $n=1$ and $n=2$
levels above shows that the $1\rightarrow0$ transition rate via \Lya scattering
is related to the total rate $P_\alpha$ by $P_{10}=4P_\alpha/27$ (Field 1958).
This relation and equation~(\ref{eq:Tspin}) are derived in the Appendix.
In the limit $T_K\gg T_{\rm CMB}$, the fractional deviation in a steady
state of the spin temperature from the temperature of the CMB is
\begin{equation}
{{T_S-T_{\rm CMB}}\over T_S}\approx \left[1+{T_{\rm CMB}\over y_\alpha
T_K}\right]^{-1}. \label{eq:ftspin}
\end{equation}
There exists then a critical value of $P_\alpha$ which, if greatly exceeded,
would drive $T_S\rightarrow T_K$. This thermalization rate is
(Madau \etal 1997)
\begin{equation}
P_{\rm th}\equiv {27A_{10}T_{\rm CMB}\over 4T_*}\approx 
(5.3\times 10^{-12} ~{\rm s}^{-1})~ \left({1+z\over 7}\right).
\label{eq:ptherm}
\end{equation}

\subsection{21-cm Emission Efficiency}

To illustrate the basic principle of the proposed observations, consider a
region of neutral material with spin temperature $T_S\neq T_{\rm CMB}$, having
angular size on the sky which is large compared to a beamwidth, and radial
velocity extent due to the Hubble expansion which is larger than the bandwidth.
Its intergalactic optical depth at $21(1+z)\,$cm along the line of sight, 
\begin{equation}
\tau(z)={3c^3 h^3 n_\nHI(0)A_{10}\over 32\pi H_0 k_{\rm B}^3 T_*^2 T_S}
(1+z)^{1.5}\approx 10^{-2.9}h_{50}^{-1}\left({T_{\rm CMB}\over T_S}\right)
\left({\Omega_{\rm IGM}h_{50}^2 \over 0.05}\right) (1+z)^{1/2},
\end{equation}
will typically be much less than unity. The experiment envisaged consists of
two measurements, separated in either angle or frequency, such that one
measurement, the fiducial, detects no line feature, either because there is no
\HI or because $T_S\approx T_{\rm CMB}$, and the second at $T_S\neq
T_{\rm CMB}$. Since the brightness temperature through the IGM is
$T_b=T_{\rm CMB} e^{-\tau}+T_S(1-e^{-\tau})$, the differential antenna
temperature observed at the Earth between this region and the CMB  will be
\begin{equation}
\delta T=(1+z)^{-1} (T_S-T_{\rm CMB}) (1-e^{-\tau})\approx (0.011~{\rm K})
h_{50}^{-1} \left({\Omega_{\rm IGM}h_{50}^2\over 0.05}\right) \left({{1+z}\over
9}\right)^{1/2} \eta, \label{eq:dT}
\end{equation}
where the 21-cm radiation efficiency is defined as
\begin{equation}
\eta\equiv x_\nHI\left({T_S-T_{\rm CMB}\over T_S}\right). \label{eq:eff}
\end{equation}
Here $x_\nHI$ refers to the neutral fraction of the hydrogen in the region
for which $T_S\neq T_{\rm CMB}$. As long as $T_S$ is much larger
than $T_{\rm CMB}$ (hence if there has been significant preheating of the
intergalactic gas), $\eta\rightarrow x_\nHI$, and the IGM can be observed in
emission at a level which is independent of the exact value of $T_S$. By
contrast, when $T_{\rm CMB}\gg T_S$ (negligible preheating), the differential
antenna temperature appears, in absorption, a factor $\sim T_{\rm CMB}/T_S$
larger than in emission, and it becomes relatively easier to detect
intergalactic neutral hydrogen (Scott \& Rees 1990).

\subsection{Preheating the IGM}

The role of the spin temperature is manifest in eq.~(\ref{eq:eff}):\ when
$T_S<T_{\rm CMB}$ the IGM absorbs 21-cm radiation from the CMB, while
for $T_S>T_{\rm CMB}$ the IGM emits 21-cm radiation in excess of the CMB.
In the absence of decoupling mechanisms, $T_S=T_{\rm CMB}$. The presence
of \Lya photons with sufficient intensity will thus enable the IGM to be
``seen.'' The adiabatic expansion of the Universe will generally bring the
kinetic temperature of the IGM well below the temperature of the CMB. Coupling
$T_S$ to $T_K$ will permit the IGM to be detectable in absorption. If there
are sources of radiation that heat the IGM, however, it may be possible instead
to detect the IGM in emission.

Possible heating sources are soft x-rays from an early generation of
QSOs or thermal bremsstrahlung emission produced by the ionized gas in
the collapsed halos of young galaxies. In CDM-dominated cosmologies, the
latter may be in sufficient number to heat the IGM above the CMB temperature
by $z\approx7$ (Madau \etal 1997).

While photons just shortward of the photoelectric edge are absorbed at
the ionization front generated by a QSO source, photons of much shorter
wavelength will be able to propagate to much greater distances. Most of
the photoelectric heating of the IGM by a QSO is accomplished by soft
x-rays. The time required for the radiation at the light front to heat
the intergalactic gas to a temperature above that of the CMB is typically
10\% of the Hubble time. The \HII region produced by a QSO will therefore be
preceded by a warming front. Note that, as the X-ray-heated bubbles around QSOs
will survive as fossils even after the quasar has died, several generations
$N_g$ of quasars may actually be responsible for preheating the entire IGM. For
a typical QSO age of $t_Q\approx3\times10^8 N_g^{-1/3}$ yr, the required QSO
comoving space density to heat the entire IGM to a temperature above that of
the CMB by $z\approx 6$ is $\sim10^{-10}\, {\rm Mpc}^{-3} N_g^{-1}$. By
comparison, the comoving space density of bright QSOs at $z=4$ is
$\sim 100N_g$ times larger (Warren, Hewett, \& Osmer 1994). If all
bright galaxies undergo a quasar phase, QSOs  must have a very short lifetime,
and $N_g\sim 100$. Soft X-rays from a few bright QSO sources could then prevent
collapsing structures, such as protoclusters while still in the linear regime,
from being detected in 21-cm absorption against the CMB.

An additional heating source is the \Lya photon scattering itself.
The average relative change in a \Lya photon's energy $E$ after having been
scattered by a hydrogen atom at rest is
\begin{equation}
\langle{\Delta E\over E}\rangle=-{h\nu_{\alpha}\over m_\nH c^2}\approx 
-10^{-8}, \label{eq:de_e}
\end{equation}
where $m_\nH$ is the mass of the hydrogen atom. (It should be noted that this
is an approximation valid only for $h\nu_\alpha\gg kT_K$. In the opposite
limit, energy will flow from the atoms to the photons.) Through recoil, energy
is transferred from photons to atoms at a rate 
\begin{equation}
\dot E_\alpha=-\langle{\Delta E\over E}\rangle h\nu_{\alpha}
P_\alpha. \label{eq:edot}
\end{equation}
where $P_\alpha$ is the \Lya scattering rate per H atom. In the case of
excitation at the thermalization rate $P_{\rm th}$, equation~(\ref{eq:edot})
becomes
\begin{equation}
\dot E_{\rm th}={{27}\over4}{{(h\nu_\alpha)^2}\over{m_\nH c^2}}
{{A_{10}T_{\rm CBR}}\over{T_*}} \approx
(220~{\rm K\,Gyr}^{-1})\left({{1+z}\over7}\right), \label{eq:Ethdot}
\end{equation}
(Madau \etal 1997).
The characteristic timescale for heating the medium above the CMB temperature
via \Lya resonant scattering at this rate is
\begin{equation}
\Delta t_{\rm heat}={2\over 9} {m_\nH c^2\nu_{10}\over h\nu_\alpha^2}
A_{10}^{-1}\approx 10^8\,{\rm yr},
\end{equation}
about 20\% of the Hubble time at $z\approx8$. The result is a finite interval
of time during which \Lya photons couple the spin temperature to the kinetic
temperature of the IGM before heating the IGM above the CMB temperature.
If \Lya sources turned on at redshifts $z_\alpha\la 10$, this interval would
present a window in redshift space near $z\approx8$ that would
enable a large fraction of intergalactic gas to be observable at
$\sim 160\,$MHz in {\it absorption} against the CMB, and so isolate the
epoch of First Light.

\section{Scenarios}

The calculations in this section were done in collaboration with Paolo Tozzi
and Piero Madau.

Two cosmological models are considered, a tilted Cold Dark Matter model (tCDM)
with $\Omega_0=1$, $H_0=50\, {\rm km s^{-1} Mpc^{-1}}$, and $\sigma_8=0.55$,
designed to match both CMB measurements on large scales and the constraint on
amplitude imposed by galaxy cluster abundances on small, and a flat open CDM
model (OCDM) with $\Omega_0=0.3$, $\Omega_\Lambda=0.7$,
$H_0=70\, {\rm km s^{-1} Mpc^{-1}}$, and $\sigma_8=1.1$, which similarly
matches both constraints. The baryon density in both models is assumed to
be $\Omega_Bh^2=0.024$, where $h=H_0/100\, {\rm km s^{-1} Mpc^{-1}}$.

We consider a scenario in which sources of \Lya photons are in sufficient
abundance throughout the universe to couple the spin temperature to the
kinetic temperature of the IGM everywhere. We further suppose the IGM has
been preheated to a temperature well above that of the CMB, either by the
same \Lya photons responsible for the coupling or by soft x-ray sources.
In this case, the IGM will emit at a rate independent of $T_S$ (the
hyperfine levels will be occupied according to their statistical weights,
$n_1/n_0=3$). Because of structure in the IGM, the emission will not be
uniform.

In Figure 1, we show the range of density fluctuations that would
be detectable in a single beam by a Square Kilometre Array
Interferometer (SKAI) at 160 MHz ($z=8$), as a function of beam size and
frequency band width, and for several assumed integration times. The
detection thresholds are scaled according to a continuum {\it rms} in a
80~MHz band at 160~MHz of 64~nJy over an 8 hour integration time
(Braun 1998; also see www.nfra.nl/skai/science/index.htm). The experiment is
based on taking differences between beams. The dashed lines show curves of
constant {\it rms} antenna temperature fluctuation within the IGM,
$\langle(\delta T_a)^2\rangle^{1/2}=\sigma_\rho\bar T_a$, where $\sigma_\rho$
is the {\it rms} relative density fluctuation of the IGM for a volume
corresponding to a given bandwidth ($\Delta\nu/\nu=\Delta z/ [1+z]$)
and angular size $\Delta\theta$. Here, $\bar T_a$ is the mean antenna
temperature from the IGM at redshifted 21-cm. At a fixed bandwidth, the antenna
temperature fluctuation increases with decreasing angular scale because
$\sigma_\rho$ increases with decreasing linear scale. Because the detected flux
is proportional to the solid angle of the beam, the detected signal decreases
with decreasing angular scale until it falls below the detection threshold,
indicated by the solid lines. The size of the detectable IGM fluctuations
differs greatly for the two models. This is because the growth of density
fluctuations ceases early on in an open universe, so that the fluctuations on
a given angular scale are much larger in the OCDM model than in tCDM at high
redshift.

\begin{figure}
\plottwo{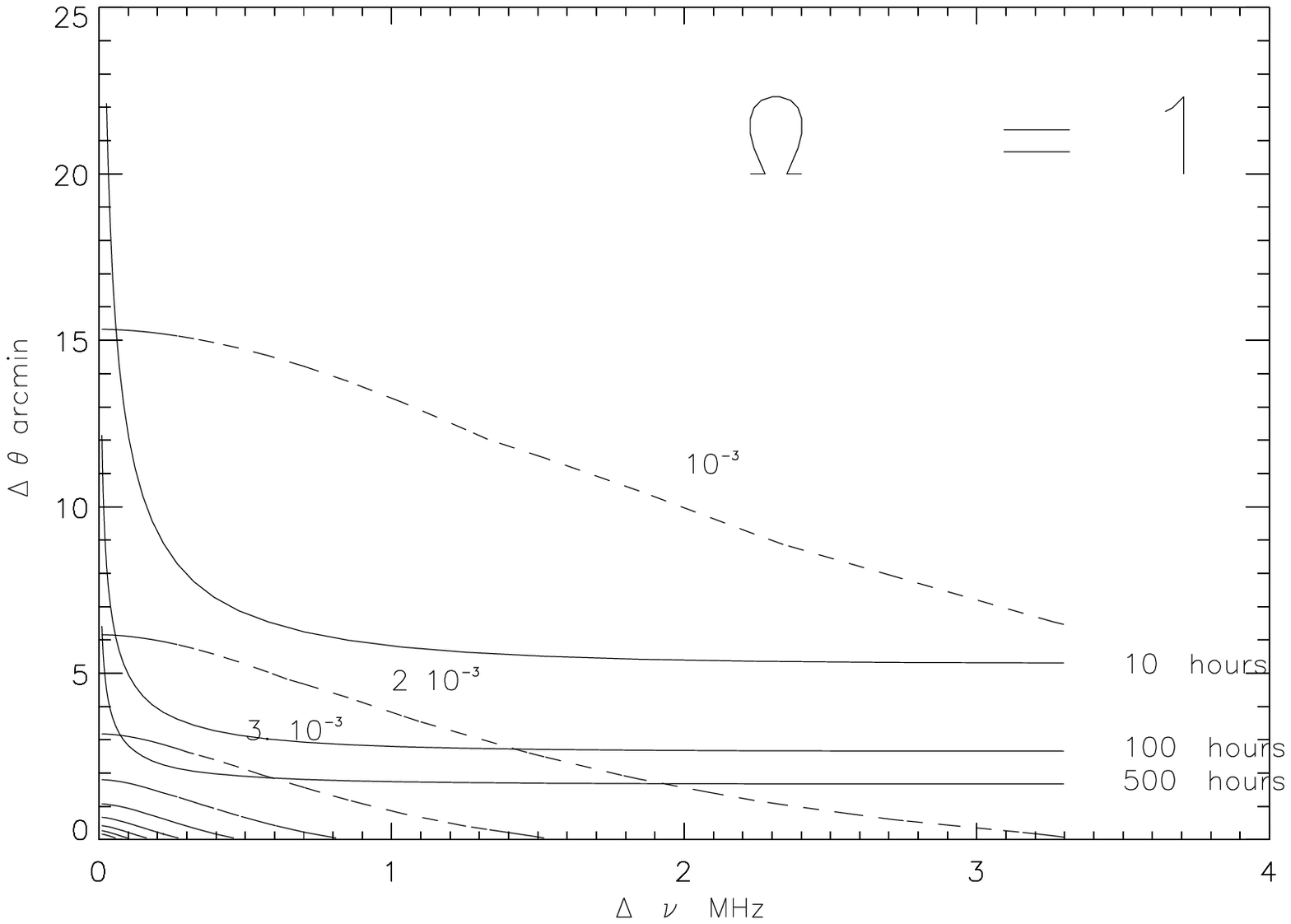}{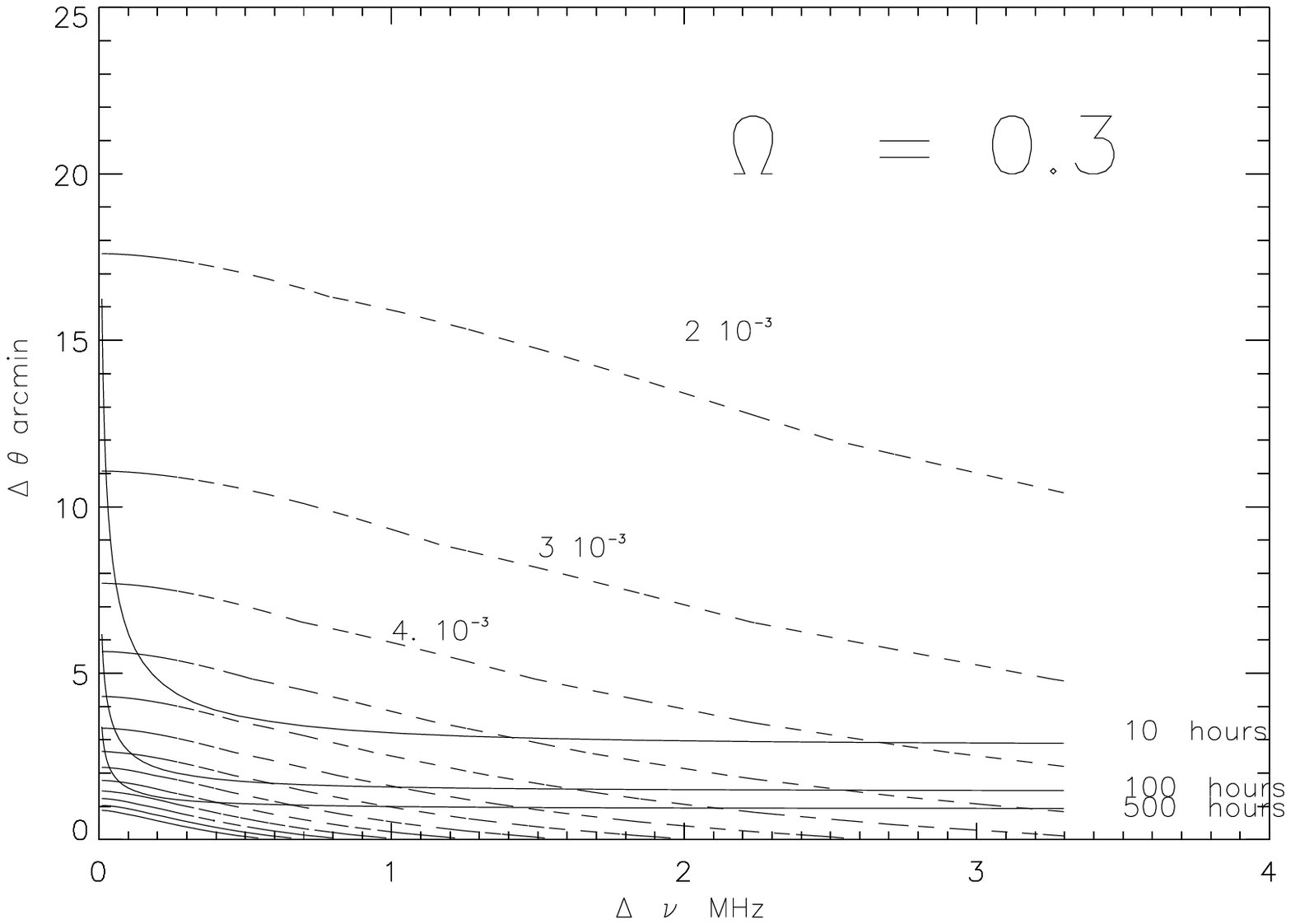}
\caption{Detectability of fluctuations in 21-cm brightness temperature
($10^{-3}<\langle(\delta T_a)^2\rangle^{1/2}<10^{-2}$), from
the IGM by SKAI as a function of beam width and bandwidth. The dashed lines
show the size of the {\it rms} fluctuations expected in a volume defined by
the angle and frequency widths. The solid lines show the threshold below which
the fluctuations are undetectable for integration times of 10, 100, and 500
hours.}
\end{figure}

In a second scenario, the spin temperature is again coupled to the IGM
kinetic temperature everywhere, but the IGM has not had time to heat above
the CMB temperature. We then consider the emission signature resulting
from a QSO soon after it turns on as the medium surrounding it is heated
by soft x-rays from the QSO. The experiment in this case is done by
differencing beams pointed in regions around the QSO with one pointed through
the QSO \HII region, where both emission and absorption are absent. An image
of the resulting emission for the tCDM model is shown in Figure 2.
The model was computed using Hydra (Couchman, Thomas, \& Pearce 1995).
As the warming front produced by the QSO expands, a growing amount of the
surrounding IGM is revealed. Note that, although the QSO was placed in the
corner of the simulation volume, the figure can equally be viewed as the
emission due to heating by a beam of soft x-rays from the QSO with an opening
angle of $90^\circ$.
Thus imaging the gas surrounding a QSO in 21-cm emission would provide a
direct means of measuring the opening angle of QSO emission.

\begin{figure}
\plotone{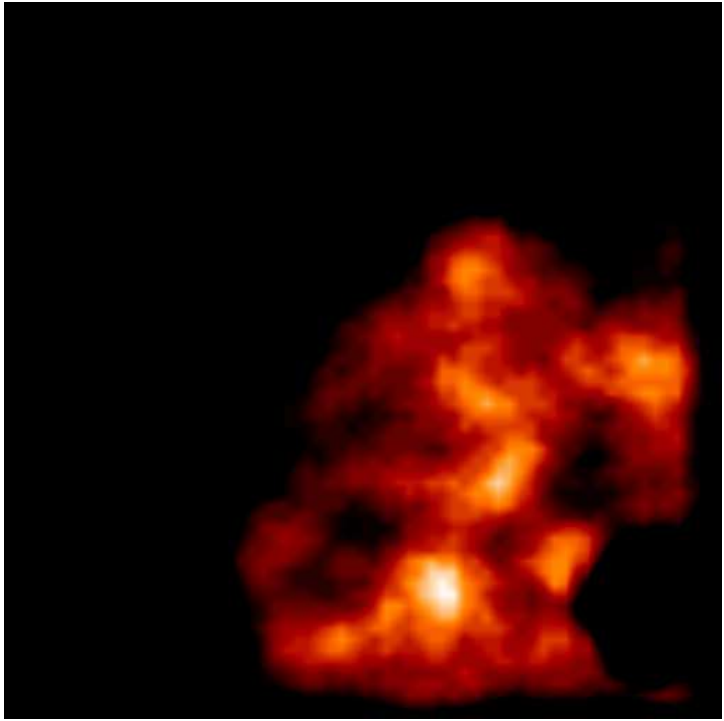}
\caption{21-cm emission from the region surrounding a QSO source (lower
right) revealed once the region is heated above the temperature of the
Cosmic Microwave Background by soft x-rays from the QSO.}
\end{figure}

It should also be noted that a region of a given fixed density
fluctuation will not always yield the same fluctuation in 21-cm emission. This
is because of the dependence of the spin temperature on the temperature of the
IGM. Only when the IGM temperature much exceeds that of the CMB will the 21-cm
emission be independent of the IGM temperature, according to
equation~(\ref{eq:eff}). In general, the fluctuations in brightness temperature
will depend on both the density fluctuations of the IGM and the
temperature fluctuations, which in turn depend on the ages and distribution
of the sources. A knowledge of the cosmological density fluctuation spectrum,
as may be measured by future CMB missions like MAP and Planck, will then enable
the statistical distribution of the sources that heat the IGM, whether QSOs as
here or \Lya photons from early stars, to be established using measurements of
the fluctuations in the 21-cm sky at high redshift.

\acknowledgments
I thank P. Madau and R. Taylor for financial support to attend the SKA
Workshop.

\appendix
\section{The Wouthuysen-Field Mechanism}

There are six hyperfine states in total that contribute to the \Lya
transition. These are shown in Figure 3. Only four of these, the two
$n=1$ states and the two $n=2$, $F=1$ states, contribute to the excitation
of the 21-cm line by \Lya scattering. Denoting the occupation number
of $_0S_{1/2}$ by $n_0$ and that of $_1S_{1/2}$ by $n_1$, the rate
equation for $n_0$ is
\begin{equation}
\frac{dn_0}{dt}=A_{10}\left(1+\frac{T_{\rm R}}{T_*}\right)n_1-
3A_{10}\frac{T_{\rm R}}{T_*}n_0
+P^\alpha_{10}n_1 - P^\alpha_{01}n_0, \label{eq:rate}
\end{equation}
where the radiation intensity at the 21-cm frequency has been expressed
in terms of the antenna temperature $T_{\rm R}$, $I_{10}=2k_{\rm B}T_{\rm R}/
\lambda_{10}^2$. The ratio $T_{\rm R}/T_*$ is the number of 21-cm photons
per mode. Here, $P^\alpha_{01}$ and $P^\alpha_{10}$ are the effective
excitation and de-excitation rates due to \Lya scattering. These may
be related to the total scattering rate of \Lya photons, $P_\alpha$,
as is now shown.

\begin{figure}
\plotone{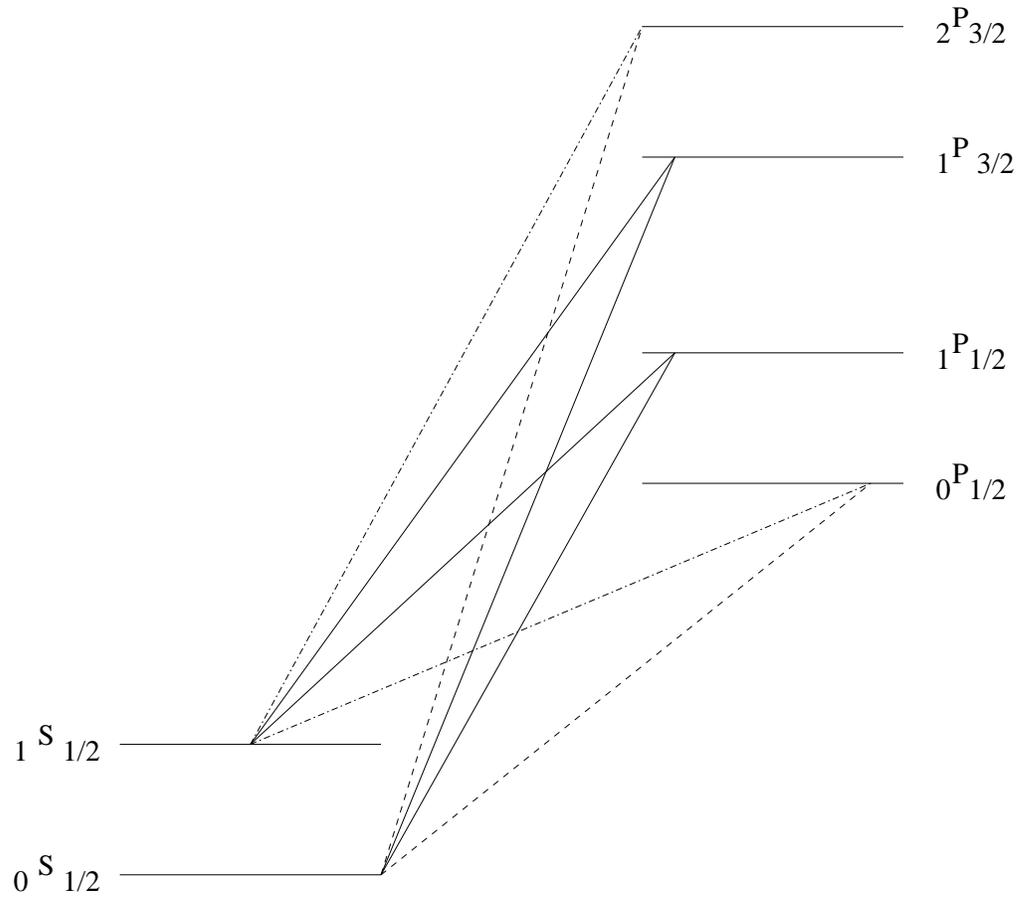}
\caption{The hyperfine energy level diagram for the \Lya transition. The
two $n=1$ states are on the left, and the $n=2$ states are on the right.
The dashed lines represent forbidden transitions ($F=0\rightarrow F=0$
and $\Delta F=2$ are not permitted), while the dot-dashed
lines represent complementary transitions that give no contribution to
the excitation or de-excitation of the 21-cm transition.}
\end{figure}

It is convenient to label the $n=2$ levels as states $2-5$, from lowest energy
to highest. Then the external radiation intensity at the frequency $\nu_{ij}$
corresponding to the $i\rightarrow j$ transition may be expressed as the
antenna temperature $T_{\rm R}^{ij}$, as above. The temperature corresponding
to the energy difference itself, $h\nu_{ij}$, is expressed as $T_{ij}=
h\nu_{ij}/k_{\rm B}$. If $A_{ji}$ denotes the spontaneous decay rate
for the transition $j\rightarrow i$, then $P^\alpha_{01}$ and $P^\alpha_{10}$
are given by
\begin{equation}
P^\alpha_{01}=3\frac{T^{\rm R}_{03}}
{T_{03}}\frac{A_{30}A_{31}}{A_{30}+A_{31}}+3\frac{T^{\rm R}_{04}}
{T_{04}}\frac{A_{40}A_{41}}{A_{40}+A_{41}}
\end{equation}
and
\begin{equation}
P^\alpha_{10}=\frac{T^{\rm R}_{13}}{T_{13}}\frac{A_{30}A_{31}}{A_{30}+A_{31}}+
\frac{T^{\rm R}_{14}}{T_{14}}\frac{A_{40}A_{41}}{A_{40}+A_{41}}.
\end{equation}

The ratios $A_{ji}/A_\alpha$, where $A_\alpha$ is the total \Lya
spontaneous decay rate, may be solved for using a sum rule for the
transitions. This states that the sum of all transitions of given $nFJ$
to all the $n'J'$ levels (summed over $F'$ and $m_{F'}$) for a given
$n'J'$ is proportional to $2F+1$.\footnote{See, e.g., Bethe, H.~A.,
Salpeter, E.~E. 1957, Quantum Mechanics of One- and Two-Electron Atoms,
New York: Academic Press.} There are
four sets of downward transitions to $n'=1$, $J'=1/2$, corresponding
to the decay intensities $I_{51}$, $I_{50} (=0)$, $I_{41}$, $I_{40}$,
$I_{31}$, $I_{30}$, $I_{21}$, and  $I_{20} (=0)$. These give
\begin{equation}
\frac{I_{51}}{I_{40}+I_{41}}=\frac{5}{3},\quad
\frac{I_{40}+I_{41}}{I_{30}+I_{31}}=1,\quad
\frac{I_{30}+I_{31}}{I_{21}}=3.
\end{equation}
Similarly, there are four sets of upward transitions to $n'=2$,
$J'=1/2$ or $3/2$, giving the intensities $I_{30}$, $I_{20} (=0)$, $I_{31}$,
$I_{21}$, $I_{50} (=0)$, $I_{40}$, $I_{51}$, and  $I_{41}$. These ratios are
\begin{equation}
\frac{I_{40}}{I_{41}+I_{51}}=\frac{1}{3},\quad
\frac{I_{30}}{I_{21}+I_{31}}=\frac{1}{3}.
\end{equation}
Using $I_{kj}/I_\alpha=(g_k/g_{\rm tot})(A_{kj}/A_\alpha)$, where $I_\alpha$ is
the total \Lya decay intensity summed over all the hyperfine transitions, and
$g_k=2F+1$ is the statistical weight of level $k$ ($g_{\rm tot}$ is the sum
of the weights of the upper levels), gives $A_{20}/A_\alpha=A_{50}/A_\alpha=0$,
$A_{21}/A_\alpha=A_{51}/A_\alpha=1$, $A_{30}/A_\alpha=A_{41}/A_\alpha=1/3$,
and $A_{31}/A_\alpha=A_{40}/A_\alpha=2/3$. We then obtain $P_\alpha=3A_\alpha
(T_{\rm R}/T_\alpha)$ ($T_\alpha\equiv h\nu_\alpha/k_{\rm B}$), and
$P_{10}=(4/27)P_\alpha$. Statistical equilibrium for $T_{\rm R}=0$ requires
$P_{01}/P_{10}=n_1/n_0=3\exp(-T_*/T_K)$. Solving the rate equation~(\ref
{eq:rate}) for $T_{\rm R}>0$ gives equation~(\ref{eq:Tspin}), where $T_{\rm R}=
T_{\rm CMB}$, and where $T_S>>T_*$, $T_K>>T_*$, and $T_{\rm R}>>T_*$ have been
assumed.

\end{document}